\documentstyle[epsfig]{mn}

\newcommand{\etal}
        {et al.}

\def\bj{b_{\scriptscriptstyle\rm J}}

\begin{document}

\title[The 2dFGRS: Galaxy luminosity functions per spectral type]
{The 2dF Galaxy Redshift Survey: Galaxy luminosity functions per spectral type}

\author[Darren S.\ Madgwick et al.\ ] {
\parbox[t]{\textwidth}{
Darren S.\ Madgwick$^1$\thanks{Email: dsm@ast.cam.ac.uk},
Ofer Lahav$^1$,
Ivan K. Baldry$^{2}$,
Carlton M. Baugh$^{3}$,
Joss Bland-Hawthorn$^4$,
Terry Bridges$^4$, 
Russell Cannon$^4$, 
Shaun Cole$^3$, 
Matthew Colless$^5$, 
Chris Collins$^6$, 
Warrick Couch$^7$, 
Gavin Dalton$^8$,
Roberto De Propris$^7$,
Simon P.\ Driver$^9$, 
George Efstathiou$^1$, 
Richard S.\ Ellis$^{10}$, 
Carlos S.\ Frenk$^3$, 
Karl Glazebrook$^{2}$, 
Carole Jackson$^5$,
Ian Lewis$^4$, 
Stuart Lumsden$^{11}$, 
Steve Maddox$^{12}$,
Peder Norberg$^3$,
John A.\ Peacock$^{13}$, 
Bruce A.\ Peterson$^5$, 
Will Sutherland$^{13}$,
Keith Taylor$^{10}$}
\vspace*{6pt} \\
$^1$Institute of Astronomy, University of Cambridge, Madingley Road,
    Cambridge CB3 0HA, UK \\
$^{2}$Department of Physics \& Astronomy, Johns Hopkins University,
       Baltimore, MD 21218-2686, USA \\
$^3$Department of Physics, South Road, Durham DH1 3LE, UK \\
$^4$Anglo-Australian Observatory, P.O.\ Box 296, Epping, NSW 2121,
    Australia\\  
$^5$Research School of Astronomy \& Astrophysics, The Australian 
    National University, Weston Creek, ACT 2611, Australia \\
$^6$Astrophysics Research Institute, Liverpool John Moores University,  
    Twelve Quays House, Birkenhead, L14 1LD, UK \\
$^7$Department of Astrophysics, University of New South Wales, Sydney, 
    NSW 2052, Australia \\
$^8$Department of Physics, Keble Road, Oxford OX3RH, UK \\
$^9$School of Physics and Astronomy, North Haugh, St Andrews, Fife, 
    KY6 9SS, UK \\
$^{10}$Department of Astronomy, Caltech, Pasadena, CA 91125, USA \\
$^{11}$Department of Physics, University of Leeds, Woodhouse Lane,
       Leeds, LS2 9JT, UK \\
$^{12}$School of Physics \& Astronomy, University of Nottingham,
       Nottingham NG7 2RD, UK \\
$^{13}$Institute for Astronomy, University of Edinburgh, Royal Observatory, 
       Blackford Hill, Edinburgh EH9 3HJ, UK \\
}

\date{
Accepted 2002 6 February.
Received 2001 23 July; 
in original form 2001 11 July}

\pagerange{\pageref{firstpage}--\pageref{lastpage}}
\pubyear{2001}

\label{firstpage}

\maketitle

\begin{abstract}
We calculate the optical $\bj$ luminosity function of the 2dF Galaxy Redshift
Survey (2dFGRS) for different subsets defined by their spectral
properties.  These spectrally selected subsets
are defined using a new parameter, $\eta$, which is a linear
combination of the first two projections derived from a Principal
Component Analysis.  This parameter $\eta$ identifies the
average emission and absorption line strength in
the galaxy rest-frame spectrum and hence is a useful indicator of the present 
star formation. We use a total of 75,000
galaxies in our calculations, chosen from a sample of high
signal-to-noise ratio, 
low redshift galaxies observed before January 2001.   We find 
that there is a systematic steepening of the faint end slope ($\alpha$)
as one moves from passive ($\alpha = -0.54$) to active
($\alpha = -1.50$) star-forming galaxies, and that there is also
a corresponding faintening of the rest-frame characteristic magnitude
$M^*-5\log_{10}(h)$ (from $-$19.6 to $-$19.2). 
We also show that the Schechter function provides a poor fit to the
quiescent (Type 1) LF for very faint galaxies ($M_{\bj}-5\log_{10}(h)$
fainter than $-$16.0), perhaps suggesting
the presence of a significant dwarf population.   
The luminosity functions presented here
give a
precise confirmation of the trends seen previously in a much smaller
preliminary 2dFGRS sample, and in other surveys.
We also present a new procedure for
determining self-consistent $K$-corrections and investigate possible
fibre-aperture biases. 
\end{abstract}

\begin{keywords}
galaxies: distances and redshifts
--
galaxies: elliptical and lenticular, cD
--
galaxies: stellar content
--
galaxies: formation
--
galaxies: evolution
\end{keywords}

\section{Introduction}
\label{section:intro}

It is well established that the measurement of the galaxy luminosity
function (LF) is sensitive to the type of galaxy being
sampled.
Morphologically early-type galaxies tend to be systematically brighter
than their late-type counterparts, resulting in LF estimates with
significantly brighter $M^*$ and shallower faint end slope $\alpha$
(e.g. Efstathiou, Ellis \&
Peterson 1988; Loveday et al. 1992; Blanton et al. 2001). 
Similar trends are also present if one selects galaxies based on
H$\alpha$ equivalent widths (Loveday, Tresse \& Maddox 1999), [OII]
equivalent widths (Ellis et al. 1996) or colour (Lin et al. 1996; Marzke \& Da 
Costa 1997). 
Understanding and being able to quantify this variation in the LF is
of great importance to a full understanding of
galaxy formation and evolution.  

Up until now the greatest obstacle
faced when attempting to determine how the LF varied with different
galaxy properties was the fact that one was required to divide already
small data sets, thereby losing much of the statistical significance
in the LF estimations.
In this paper we address this issue by making use of a subset
of the galaxies
observed to date in the 2dF Galaxy Redshift Survey
(2dFGRS Maddox 1998; Colless 2001).  This subset is the largest single
data set used in the calculation of LFs and can easily be divided
several times whilst still maintaining very precise statistics.

The 2dFGRS is a joint
UK-Australian effort to map the distribution of galaxies down to an extinction
corrected $\bj$ magnitude of $19.45$  (median redshift $z\simeq0.1$).
  In so doing we expect to
obtain 250,000 galaxy spectra from which redshifts can be determined.
This is a factor of 10 more than any previous redshift survey.

In order to divide our data set in a meaningful way we develop in this
paper a classification of the galaxies based upon their
observed spectra.
As with other astronomical data, there are different approaches for
analysing galaxy spectra (see e.g. Lahav 2000 and references therein). 
If one has a well defined physical model for galaxy spectra, then 
it is appropriate to estimate parameters of interest 
(such as age and star-formation rate) e.g. by Maximum Likelihood,
directly using all the spectral bins, or via a compressed version of the data 
designed to give maximum information on the physical parameters of interest
(e.g. Heavens, Jimenez \& Lahav 2000).
If, on the other hand, one prefers to let the data `speak for themselves',
in a model-independent way, then it is more useful to look at the 
distribution of the galaxies in the high-dimensional space defined
by the spectral bins.  It is then possible to either look for distinct 
groups, e.g. early and late types, or more refined classes
(e.g. Slonim et al. 2000). An alternative, which we present here, 
is to find a continuous (sequence-like) parameterisation of the spectral
features, which can later be divided into subsets. 
The parameterisation we develop is denoted by $\eta$ and essentially
represents a measure of the average absorption or emission line
strength present in each galaxy's spectrum.  This classification is
robust to the known instrumental uncertainties and also has the
advantage of being easily interpreted in terms of the current
star-formation present in each galaxy.

The set of galaxies we use in this paper comprises a
sample of 75,000 galaxy spectra.  This is more than 12 times
larger than the one used in our previous analysis (Folkes et al. 1999,
hereafter F99) and is by far the largest sample used to date.

As well as showing 
the latest LF determinations we will also present a new procedure for
calculating self-consistent $K$-corrections and investigate
possible fibre-aperture biases. 

The outline of this paper is as follows. 
Section~\ref{section:data} briefly summarises the 2dFGRS and describes the
data set we are using, Section~\ref{section:spectypes} outlines our method for dividing the
data set based on our parameterisation of the galaxy spectral type, $\eta$.
Section~\ref{section:lumfunctions} gives a detailed description of the calculations involved in
estimating the LF and in Section~\ref{section:discussion} we discuss our results and future work.

\section{The 2\lowercase{d}FGRS data}
\label{section:data}

\begin{figure*}
 \epsfig{file=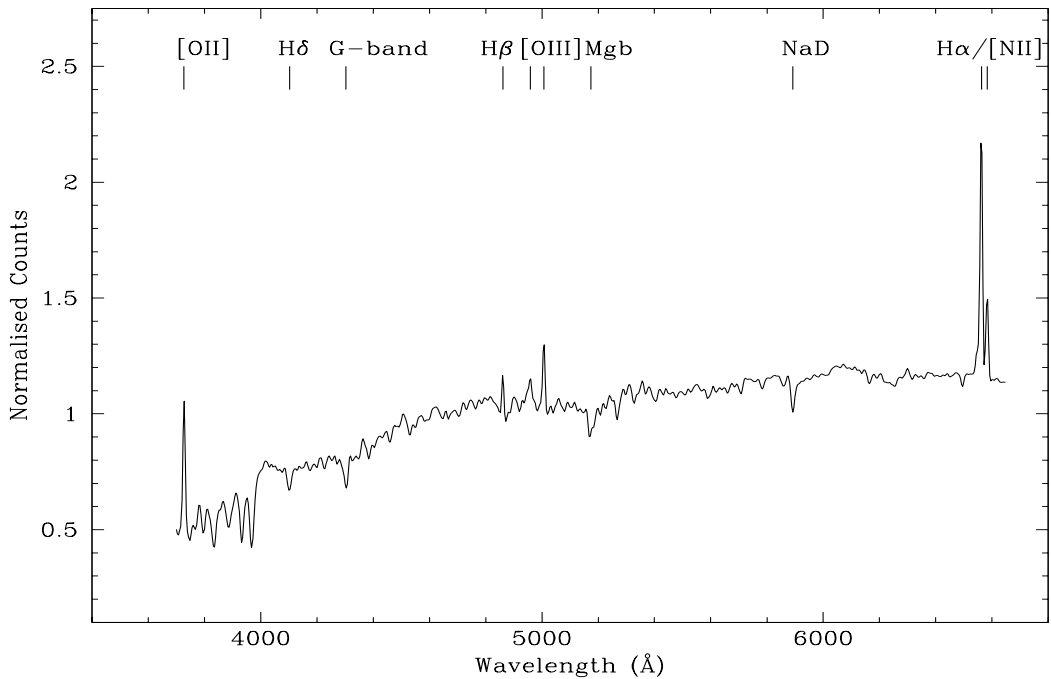,width=6in}
 \caption{The average spectrum of the $M_{\bj}-5\log_{10}(h)>-18$
 volume-limited sample of galaxies.  The main spectral features
 present are labelled.}
 \label{aver}
\end{figure*}

The 2dFGRS has already observed
approximately 200,000 unique galaxies for which it has obtained
redshifts. 
The survey, once complete, will cover approximately
2000$\sq\degr$ on the sky, split between two independent strips: one in
the northern Galactic hemisphere and the other centred roughly on
the south Galactic pole.  In addition to this there are 99 random fields in 
the southern Galactic cap.

The 2dF instrument itself is capable of observing up to 400 galaxy
spectra simultaneously (Lewis et al., 2001).  Each
galaxy has been   
selected from a revised and extended version of the Automated Plate Measuring 
(APM) galaxy
catalogue (Maddox et al. 1990) in order to determine its
position and magnitude.  It is then observed as part of the survey
through a 140$\mu$m ($\sim2$ arcsec) diameter optical
fibre.  The $b_{\rm J}$ magnitudes we use in this analysis
are 
total magnitudes derived from updated versions of the original APM
scans.  These magnitudes have been updated to take into account new
CCD calibration data (see Colless et al. 2001 and Norberg et al.,
2001)  and are believed to have an rms error 
of approximately $\pm0.15$ magnitudes.

Each of the spectra observed using the 2dF instrument spans 1024
channels with a spectral scale of 4.3$\rm{\AA}$ per pixel; the FWHM is
measured by arc lines to be of the order of 1.8-2.5 pixels.  Typically
at the survey limit the observed spectra have an
average signal-to-noise ratio of
$\sim10$ per pixel, sufficient for determining redshifts and
performing spectral 
analyses.  For the purposes of our spectral analysis we have
restricted ourselves to the redshift range of 
$0.01<z<0.2$ so that we are left with a uniform sample of galaxy
spectra in the 
rest-frame wavelength range spanning [OII] to H$\alpha$.  A detailed
outline of the spectral reduction pipeline is given in Colless et
al. (2001) and F99 which we follow
with only minor modifications.  

The data that we have available for this analysis include all
observations up until January 2001.  
At this stage a total of 111404 and 60062 galaxy spectra had been
obtained in the SGP and NGP regions respectively.  From these objects
we remove those which did not receive accurate redshifts ($Q<3$,
Colless et al. 2001),
repeated observations and spectra with particularly low
signal-to-noise ratio ($\le10$).  This leaves us 
with 78994 (SGP) and 44937 (NGP) galaxies for use in our spectral analysis.
Because of interference from sky emission and atmospheric absorption we
have made a further restriction to our redshift range such that
$z\le0.15$.  This cut ensures that the H$\alpha$ line in our
spectra is not corrupted.  Note that this is only a problem for
the small redshift range $0.15 < z < 0.17$, however, we have
also excluded the galaxies with $z>0.17$ in order to simplify our
analysis.  Imposing this
cut and removing spectra observed in poor quality fields (Section~4.1)
leaves us with 43449 (SGP) and 32140 (NGP) galaxies. 
A total of 75589 unique galaxies will therefore be used in our
subsequent analysis and measurement of the LF.

\section{Spectral types}
\label{section:spectypes}

The spectral classification presented here is based upon a Principal
Component Analysis (PCA) of the galaxy spectra.  PCA is a statistical
technique which has
been used with considerable success by multiple authors in the past
(e.g. Connolly et al. 1995; Folkes, Lahav \& Maddox 1996; Glazebrook,
Offer \& Deeley 1998; Bromley et al. 1998; F99) to deal with large
multi-dimensional data sets. 
A detailed mathematical formulation of the PCA adopted here in
is given in F99.  Note that the most significant difference
between this 
formulation and that used by other authors (e.g. Connolly 1995) is that our
spectra have 
been mean-subtracted (Fig.~\ref{aver}) before constructing the
covariance matrix. This makes no substantial
difference to the analysis since using other methods simply yields the
mean spectrum as the first component.  

We note that throughout the remainder of this paper we will denote 
the eigenvectors (herein eigenspectra) as
$\bmath{PC_1}$,$\bmath{PC_2}$ etc. and the projections 
onto these axes by $pc_1$, $pc_2$ etc.

\subsection{Applying PCA}

PCA is a useful technique in that it allows us to easily visualise  
a multi-dimensional population in terms of just a handful of
significant components.  It does this by identifying the components of 
the data (in this case the galaxy spectra) which are the most discriminatory
between each galaxy.  The
significance of each component is measured in terms of its
contribution to the variance over the sample and is determined
in the PCA.  This allows us to identify just the most
significant components for future use. 
It is clear from such a formalism that any clustering in a space
defined by the PCA is indicative of distinct sub-populations within the
sample 

In terms of reduced dimensionality we find, after applying
the PCA to our galaxy spectra, that
rather than using the original 738 spectral
channels to describe each spectrum we can use only two projections
and still retain two thirds of the total variance within
the population.  Indeed we can retain 50\% of the total variance just
from the first component (the total variance of the population
includes noise 
and hence this is an underestimate of the true fraction of variance in
these projections).   The significance of each successive
principal component drops off very sharply so these first few
components are by far the most important (Table~\ref{tabpc}).

\begin{table}
 \caption{The relative importance (measured as variance) of the first
    8 principal components.}
 \begin{tabular}{@{}cccc@{}}
   \hline
   Component & Variance (\%) & Component & Variance (\%)\\
   \hline
     1 & 51 & 5 & 1.7 \\
     2 & 15 & 6 & 1.3 \\
     3 & 3.4 & 7 & 0.99 \\
     4 & 2.4 & 8 & 0.84 \\
   \hline
 \end{tabular}
 \label{tabpc}
\end{table}

Admittedly it is not clear a priori whether retaining maximum
variance over the spectral population also retains a corresponding
maximum of physical information regarding each galaxy. However, in this
paper, and also in previous independent analyses, it is found
empirically that the components dominating the spectral 
distribution do in fact relate well to physical attributes 
For example, morphology and other well 
known diagnostics such as the continuum slope (colour), the average
emission/absorption line strength and the strength of the H$\alpha$
line (see e.g. Ronen, Arag\'on-Salamanca \& Lahav 1999).

It is worth noting here that there are several limitations to using PCA
which the reader should bear in mind.  The most important of these is
that it imposes linearity on the data set. This is certainly not a
satisfactory approximation in all cases, however the general
ability of the analysis to successfully reconstruct galaxy spectra
with only the first few projections suggests that such an
approximation is valid.
We also note that  
in this analysis (as in previous work) we have restricted ourselves to
deriving only the most general classification scheme.  In so doing we
ignore several distinct sub-populations in our data set such as
star-bursts and
AGNs.  In fact it should be possible to take another approach and to
target the analysis on finding these sub-populations.

\subsection{Volume-limited PCA}

Due to the large number of galaxy spectra acquired so far in the 2dFGRS it
has now become possible to consider volume-limited subsets of the galaxies
for use in the PCA.  This approach is much more optimal
compared to considering the entire flux-limited data set in which the
spectral population can 
be biased by relationships between the luminosity and type of
galaxies.  

\begin{figure}
 \epsfig{file=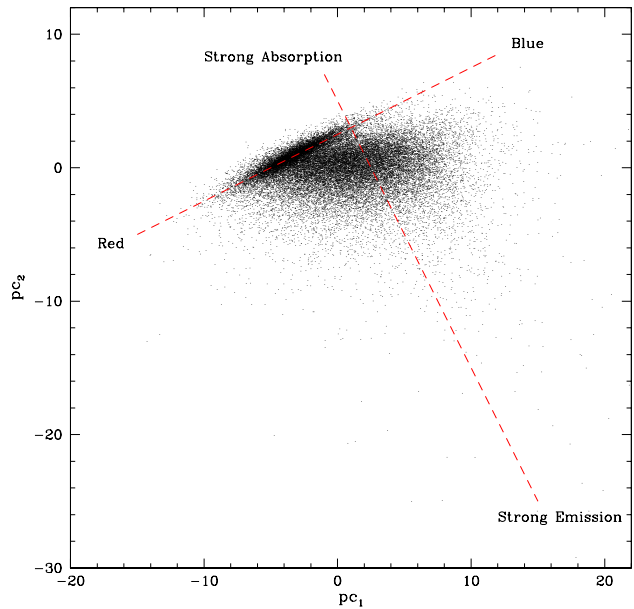,width=3.5in}
 \caption{Projections ($pc_1$ and $pc_2$) in the space defined by the
 first two eigenspectra derived from the PCA for the
 $M_{\bj}-5\log_{10}(h)<-18$ sample of galaxy spectra.
 The trends illustrated have been 
 derived from the eigenspectra by identifying the linear combinations
 that either maximise the influence of the absorption/emission features
 or minimise them.}
 \label{proj}
\end{figure}

We find that using an absolute $M_{\bj}-5\log_{10}(h)$ magnitude limit
of $-$18.0 
(corresponding to a redshift of $z\simeq0.1$) gives a representative
subset of the local population whilst still retaining 35,289
galaxies in the analysis, a very large fraction of the total.  
We have therefore chosen to perform the
PCA on this subset of the population in order to determine the principal
components.  It is then a trivial matter to use these components to
derive the individual projections for each of the 
remaining galaxies outside of our volume limited sample. 
The average spectrum of this volume-limited sample is shown in
Fig.~\ref{aver} and the projections in the plane defined by the first
two principal components are shown in Fig.~\ref{proj}.

\subsection{Interpreting the PCA}

\begin{figure*}
 \epsfig{file=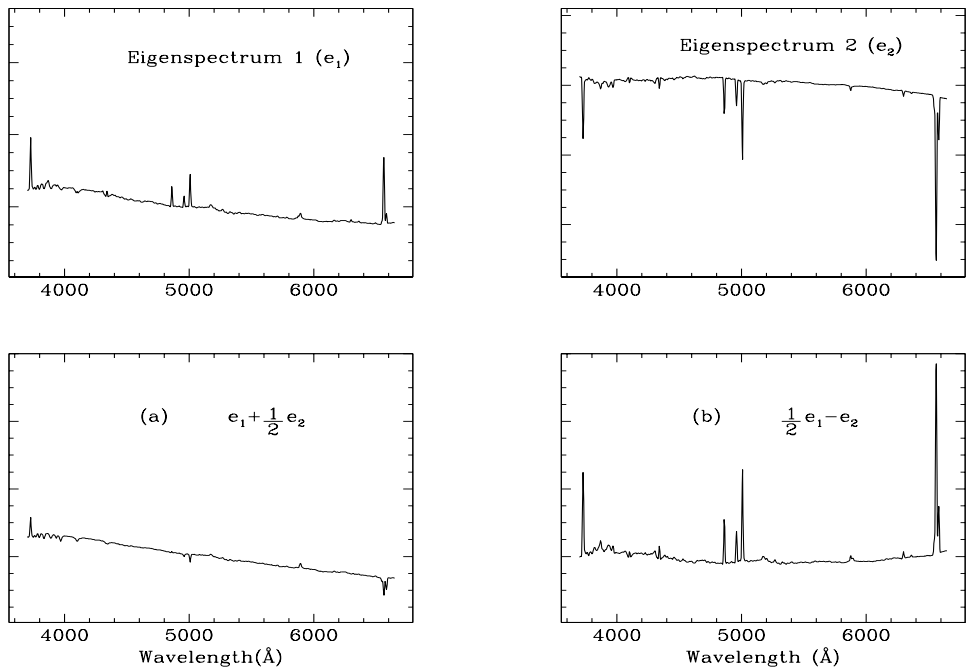,width=6in}
 \caption{The first two eigenspectra are shown in the top two panels
 and the linear combinations which either minimise (a) or maximise (b)
 the effect of emission/absorption features are shown in the bottom two.}
 \label{cevec}
\end{figure*}

We can identify general trends in the distribution of projections
shown in Fig.~\ref{proj} by 
considering the physical significance of the first two eigenspectra. 
We find 
that whilst the first eigenspectrum contains an approximately
equal contribution from both the continuum
and the emission/absorption line strengths, the
second is dominated by the latter.
Therefore
by taking certain linear combinations of the two we can either
maximise or minimise the contribution from the line features in the
spectrum (Fig.~\ref{cevec}).  Such a transformation of axes allows us
to easily interpret the distribution of Fig.~\ref{proj} as both a
sequence in colour from red (bottom left) to blue (top right) 
and an orthogonal
sequence from absorption (top left) to emission (bottom right).

\subsection{Instrumental effects}

The 2dF instrument was designed to measure large numbers of redshifts
in as short an observing time as possible.  In so doing it has made very
large projects
such as the 2dFGRS possible. However, in order to optimise the number of
redshifts that can be measured in a given period of time, compromises
have had to be made with respect to the spectral quality of the
observations.  Therefore if one wishes to characterise the observed
galaxy population in terms of their spectral properties care must be
taken in order to ensure that these properties are robust to the
instrumental uncertainties (see e.g. Lewis et al., 2001, for a
more detailed discussion).  

The 2dF instrument makes use of up to 400 optical fibres with a
diameter of 140$\mu$m (corresponding to $2.0 - 2.16''$ on the sky, depending on
plate position).  The quality and representativeness of the observed
spectra can be compromised in many ways, the most significant of which
are:
\begin{itemize}
\item Astrometric errors in determining the actual position of the
galaxy on the sky can result in the fibre being placed in a position
that is not optimal for light collection.  In general the position of a
galaxy centre is known to within $0.4''$.  
\item Given a known galaxy location on the sky the robotic positioner
will place the fibre to within 20$\mu$m.  The RMS error in placing
this fibre is found to be 11$\mu$m ($0.16''$).
\item The observed colour (continuum slope) of the galaxy depends on
the positioning of the fibre aperture.  This is due to a feature of the
2dF corrector lens design which results in 
relative spatial displacements between different components of a
galaxy's spectrum.  These displacements can be as
large as $1''$ (at the $20'$ radius) between the incident blue and red
components of the observed spectrum.  This together with the previous
two uncertainties in placing the fibre aperture will result in
uncertainties in the  continuum measurement which make the spectra
very difficult to accurately flux calibrate.
\item The apparent size of a galaxy on the sky can be larger than
the fibre aperture.  The positioning  
of this aperture can therefore result in an unrepresentative spectrum
due to spectral gradients within the galaxy.
\item Atmospheric dispersion (due to imperfect correction) and the
differential atmospheric refraction during an exposure each result in
errors of the order of $<0.3''$.
\end{itemize}

The source of the chromatic dispersion mentioned above is due to
a design feature of the 2dF corrector lens which gives us our extended (2
degree) field of view.  This dispersion, combined with the random
positioning errors, results in uncertainties in calibrating
the continuum of each galaxy spectrum.  For this reason we will not
make use of the information contained in the continuum of each galaxy
in our subsequent analysis.
However, we will make use of the continuum calculated by
averaging over a sufficiently large ensemble of galaxy spectra since
this will be much more robust (see Section 4.2 for further details).

The random errors in positioning the fibre aperture could also result
in uncertainties in measuring the line-strengths of emission features
such as H$\alpha$ since we would expect the strength of this emission
to vary spatially across the galaxy.  However, because the random positioning
errors are much 
smaller than the scale lengths of the galaxies this isn't a large
problem.  For this reason the information contained in the
emission/absorption line strengths should be relatively robust.

Perhaps our greatest concern regarding the representativeness of the
spectra is due to the limited size of the aperture with respect to the
galaxy.  Surprisingly, we have found it difficult to isolate any
systematic bias from this effect.  Of course this effect will be substantially
diluted in the observed spectra due to the significant seeing present
at the Anglo-Australian Telescope in Siding Springs, which is of the
order of $1.5''-1.8''$.  In addition the presence of differential
atmospheric refraction may also assist us.
We discuss the impact of this `fibre-aperture bias' on our results in
more detail in Section 5.3.

\subsection{$\eta$ parameterisation}

We are presented with several options in attempting to derive a
classification for the observed 2dF galaxies based upon their
spectra.  However, there are several issues that one must first consider:
\begin{itemize}
\item The distribution of galaxies in the ($pc_1$,$pc_2$) plane is
smooth (although slightly bimodal), representing a continuous sequence
from absorption to emission 
and from red to blue.  There is no evidence to suggest any division into
distinct classes of galaxy spectra.
\item The instrumental errors result in unstable
continuum measurements. On the other hand small scale features such as
line-strengths are relatively unaffected.
\item The relationship between morphology and spectral properties is
not very well understood and hence attempting to anchor the
classification using a 
morphological training set (e.g. Connolly et al. 1995; F99) would be
premature.  A 
more robust and quantifiable measure of the galaxy type is
required.
\end{itemize}
By projecting the $pc_1$ and $pc_2$ components of each galaxy onto the
linear combination which maximises the effect of emission/absorption
line features we are in effect high-pass filtering the spectra.  Thus we
would expect (and indeed we find) this projection to be relatively stable
to uncertainties in continuum measurements.

\begin{figure}
 \epsfig{file=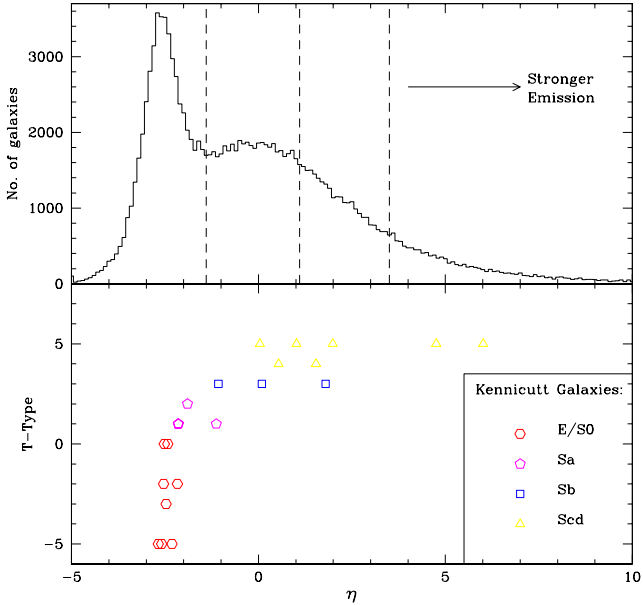,width=3.5in}
 \caption{The observed distribution of spectral type measured by
 $\eta$.  Also shown
 are the four divisions we use to
 divide the dataset (see Fig.~\ref{specfig} for mean spectra).  The
 bottom panel shows the 
 correlation between $\eta$ and morphological type using a training set
 of galaxies taken from the Kennicutt Atlas (Kennicutt 1992).}
 \label{eta}
\end{figure}

By using this projection we are determining a measure of the average
emission/absorption line strength of a galaxy which is easily quantifiable
and robust. In addition this projection is also representative of the
spectral sequence of the galaxy population since it is composed of
the two most significant principal components (representing 66\% of
the total variance over the population). 

We therefore choose to adopt this projection, which we shall denote
$\eta$, as our continuous measure of spectral type
\begin{equation}
 \eta = a\; pc_1 - pc_2 \;.
\end{equation}
The value of $a$ which maximises the emission/absorption features is
essentially identical to that we would find from identifying the most
stable projection in the ($pc_1$,$pc_2$) plane between repeated pairs
of spectral observations.  Using this method we find $a = 0.5 \pm 0.1$.  

In Fig.~\ref{eta} the distribution of the $\eta$ projections is
shown for the galaxies observed to date in the 2dFGRS. Also shown in
the same figure is
the $\eta$-morphology relation for a sample of galaxies from the
Kennicutt Atlas (Kennicutt 1992).  Comparing the two data sets shows that
there is a correspondence between the sequence of $\eta$ and that of
morphology. Note that this correspondence can only be treated as an
approximation since our sample of Kennicutt galaxies is not complete
and only represents a few high signal-to-noise ratio spectra with well determined
morphologies.  However, the trend is clear.

\begin{figure}
 \epsfig{file=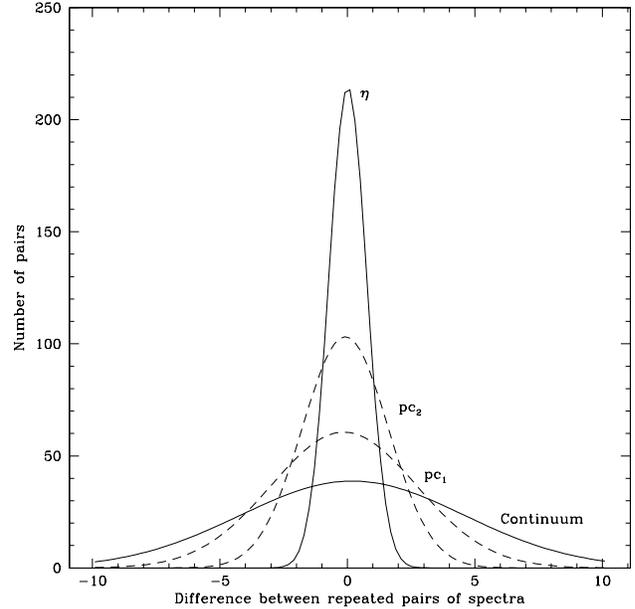,width=3.5in,height=3.5in}
 \caption{Dispersion in the PCA projections ($pc_1$ and $pc_2$) and in
 the spectral classification $\eta$.  These distributions are well fit
 by Gaussians (shown here) and were
 calculated by taking the differences in these quantities between $\sim 2000$
 repeated spectral observations.  Also shown is the dispersion in the
 orthogonal projection to $\eta$, which quantifies the continuum slope
 in the spectra.}
 \label{errors}
\end{figure}

It is important to have an understanding of the uncertainties in the
principal component analysis as this will affect the reliability of
our classification.  Using a sample of ($\sim 2000$) repeated spectral
observations we can attempt to quantify these uncertainties.
The differences in the PCA projections ($pc_1$ and $pc_2$) and in the
spectral classification between repeated pairs is shown
in Fig.~\ref{errors}.  Note that for clarity we show here the Gaussian
fits made to these error distributions.  It is clear from this figure
that the uncertainty in the $pc_1$ and $pc_2$ projections is
substantial ($1\sigma$ dispersions of 2.9 and 1.7 respectively),
however, this dispersion is greatly reduced by using the 
$\eta$ parameterisation ($1\sigma=0.7$).

\begin{figure}
 \epsfig{file=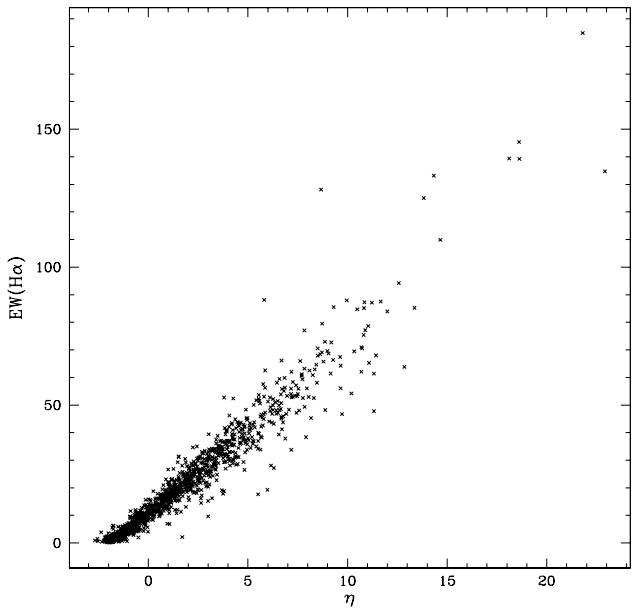,width=3.5in,height=3.5in}
 \caption{The EW(H$\alpha$) is very tightly correlated to $\eta$ for
 emission line galaxies (see
 Bland-Hawthorn et al., in preparation, for further details).}
 \label{halpha}
\end{figure}

In addition to correlating with morphology, $\eta$ has also been shown
to be strongly correlated with the equivalent width of H$\alpha$ in
emission line galaxies, Fig.~\ref{halpha} (Bland-Hawthorn et al., in
preparation). Hence $\eta$ can be
interpreted as a measure of the current star-formation present in each
galaxy.  

\begin{figure*}
 \epsfig{file=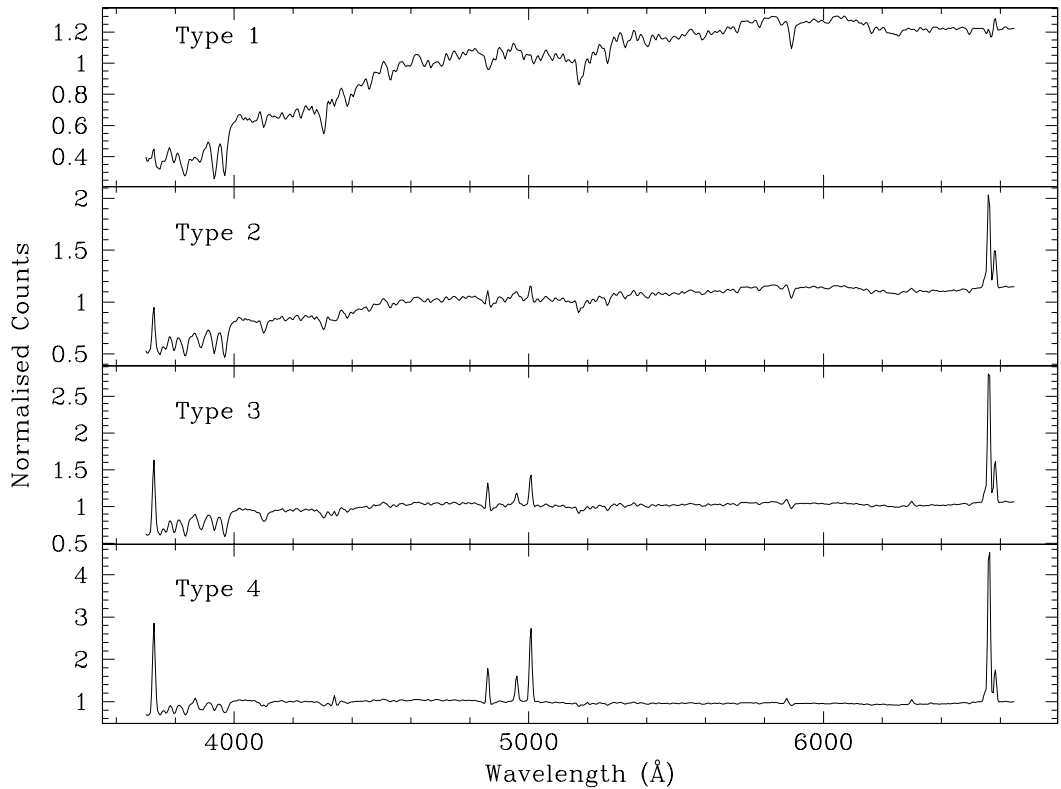,width=6in}
 \caption{The average 2dFGRS spectrum of each spectral type is shown.
 Each spectrum is in units of counts per pixel and has been normalised to have
 average counts of unity.}
 \label{specfig}
\end{figure*}

Note that having a continuous measure of spectral type
allows us to take greater advantage of having such a large data set in
which much of the detail in the spectral sequence would be smeared out
by rough divisions (as in the case of morphological segregation).
However, for the purposes of this preliminary analysis we choose to
split the spectral sequence into four broad bins (`types') as shown in
Fig.~\ref{eta}.
The divisions we have made are not determined by the morphological
training set (as in previous work) but rather by the 
shape of the $\eta$-distribution itself.
The peak in the distribution arises from the
degeneracy in spectral line-features between elliptical and early-type
spiral galaxies.  We separate this peak from the shoulder, which we
then divide into two.  Another cut is then made to separate the tail of
the distribution which will be dominated by particularly active
galaxies such as starbursts and AGN.  The
average spectrum of each type is shown in Fig.~\ref{specfig}.

\section[]{Luminosity functions}
\label{section:lumfunctions}

\subsection[] {Completeness}

We find that at a depth of $\bj=19.2$ the density of galaxies is
$149.4/\sq\degr$ over our targeted area of the sky (Norberg et
al., 2001).  In order to avoid fibre 
collisions it is not possible to assign a fibre to every galaxy in a
given field.  We take into account this configuration completeness
(relative to the parent catalogue) of $93\%$ to calculate the
effective area of the observed galaxies, thereby providing the overall
normalisation for our LF estimates.

In addition to the configuration completeness it is also necessary
to correct for the classification incompleteness: we chose to exclude
spectra with $Q<3$ and signal-to-noise ratio $\le10$, which do not yield an
accurate value of $\eta$. 
We assume that this completeness can be parameterised only
in terms of the apparent $\bj$ magnitude and define
\begin{equation}
C(\bj) = \frac{\rm{\# Galaxies\,\, that\,\, could\,\, be\,\,
classified}}{\rm{\# Galaxies\,\, targetted}} \;.
\end{equation}
This ratio is independent of redshift 
and is appropriate
for applications based on conditional probabilities $p(M|z)$ e.g. the
STY LF estimator (Section~4.3).
In the case of applications where the redshift limits are explicitly set
this completeness ratio is 
adjusted to take account of this, although this is only a small correction.
We find that the classification completeness can be well fit by a
function of the form,
\begin{equation}
C(\bj) = \alpha - \exp[\beta(\bj - \gamma)] \;.
\end{equation} 
We determine the best fit parameters by weighting each bin according
to the number of galaxies that contributed to it and performing a
non-linear least squares fit, yielding
($\alpha$,$\beta$,$\gamma$)=(0.96,1.51,20.73).  A weight for each
galaxy can then be determined from the inverse of this quantity.

In addition we note that this parameterisation can be significantly
skewed in observed fields with low overall completeness.  So before
calculating these parameters we have imposed a
minimum completeness threshold for each observed field of $90\%$.
This additional cut reduces the number of galaxies to be used in our
calculations by approximately 8,000 (as described in Section~2).

The completeness of our sample may also be influenced by other
selection biases which vary systematically with redshift or spectral
type e.g. surface brightness.  Of course it is not possible to
determine a simple correction for these effects since we do not know
the redshift or spectral type of a galaxy until it has been
successfully observed.  We expect these additional effects to be
small, they will be discussed in more detail in a forthcoming paper (Cross et
al., 2001).

\subsection{$K$-Corrections}

Here we develop a self-consistent method for $K$-corrections based on the observed 2dF
spectra.  This is a significant improvement compared with previous
analyses  
where the $K$-corrections were anchored to standard SEDs (e.g. Pence 1976)
through the relation between morphology and spectral type. 

The main obstacle in calculating $K$-corrections directly from the
observed spectra
is the large uncertainty associated with measuring the continuum 
(Section 3.4).  If this were not an issue then the calculation 
of the $K$-corrections would be straightforward, since
both the rest-frame and observed wavelength range of the 2dFGRS
spectra span the range of the $\bj$ passband where the transmission is above
5\% of the passband peak (3710 $-$ 5520$\rm{\AA}$).

On a galaxy-by-galaxy basis the continuum uncertainties caused by
the chromatic distortion in the telescope optics 
coupled with the small random fibre positioning errors will induce
random errors in the $K$-corrections.  We are assisted by the fact that
there are only three substantial
systematic effects which might affect our calculations:
\begin{enumerate}
\item Measuring the $\bj$ passband.  (Hewett \& Warren, private communication).  
\item Determining the 2dF system response (Lewis et al., 2001).
\item Aperture bias (Sections 3.4 and 5.3) due to the relatively small size of
our fibre 
aperture compared with the spatial extent of low-redshift galaxies.
\end{enumerate}
Of these three the 2dF system response is by far the least well
determined.  The system response 
is calculated using photometric standard star observations with the
2dF instrument setup and accounts for the relative photon capturing
efficiency of the 
instrument over different wavelength bands (corrected for seeing).
The exact nature of this response is still in the process of being
fully determined. 
For the mean-time we adopt a small correction to the previously
measured standard-star system response in order to correct the average
galaxy survey observation, 1997-2001, (Baldry et al., 2001).

The aperture bias mentioned in the third point can have quite a
significant impact on the observed 2dFGRS spectra since for low-redshift
galaxies we will systematically 
sample more of the (redder) inner bulge light in our spectra.
However, this will not lead to significant errors in the absolute
luminosities since at low 
redshifts, where the aperture bias can be substantial, the $K$-corrections
are small in absolute terms; conversely, at higher redshifts where the
$K$-corrections are large, the aperture bias is minimal. 
We have conducted several tests to estimate the significance of
aperture bias
on our spectra and we have found it to be negligible beyond $z\sim
0.1$, particularly due to the relatively poor seeing at the AAT which 
effectively `smears' out spectral gradients in each galaxy.  In
addition we 
find that it will only have a significant impact for particularly
low-redshift 
galaxies.  

Each individual galaxy spectrum has large uncertainties associated with
its continuum, so to calculate meaningful $K$-corrections we 
adopt here a practical approach.  Having divided the
galaxies into their respective classes we proceed to calculate the
average spectrum for each class.  Since each class contains many
thousands of galaxies it is reasonable to expect any unphysical
features present in the spectra to average out (assuming we have
accounted for the above systematic uncertainties). 
We find from repeated spectral observations that it is equally likely
for a continuum to be bent one way as the other, so
assuming we have averaged over a sufficient number of galaxy spectra
it is reasonable to assume that these effects will balance out.
Therefore each average spectrum can be considered to be a
characteristic galaxy template that reflects the true average SED of
that ensemble of galaxies.  We can now proceed to calculate
$K$-corrections for each of these four templates.  

Note that we
apply these $K$-corrections to each galaxy individually to calculate its
absolute magnitude.  However, on a galaxy-by-galaxy basis these
derived absolute magnitudes will not be optimal.  Rather, they should be 
interpreted as
part of the ensemble to which they belong.  We should therefore only
apply these $K$-corrections to applications which involve comparing
ensemble properties (e.g. LFs) or alternatively to calculating a
property of the entire data set.

We have fit quadratic forms to the $\bj$ $K$-corrections given by the
following expressions;
\begin{equation}
k_1(z) = 2.6z + 4.3z^2
\end{equation}
\begin{equation}
k_2(z) = 1.9z + 2.2z^2
\end{equation}
\begin{equation}
k_3(z) = 1.3z + 2.0z^2
\end{equation}
\begin{equation}
k_4(z) = 0.9z + 2.3z^2 \;.
\end{equation}
Where $k_i$ refers to the $K$-correction of type $i$ and is valid over
the redshift range $0<z<0.2$.
The average $K$-correction over all spectrally classified galaxies is
given by $k_{\rm av}(z) = 1.9z+2.7z^2$.  This expression can be used in
lieu of the above or instead could be applied to non-classified
galaxies which require $K$-corrections. 

The most substantial source of errors in our $K$-corrections will
arise from the uncertainty in the 2dF system response. Taking this
into account we estimate the uncertainty in our $K$-corrections to be 
of the order of $20\%$.

Norberg \etal\ (2001) have also calculated $K$-corrections for the
2dFGRS, independently of the observed galaxy spectra. They find very
similar $K$-corrections, which are consistent with our own to within
the stated uncertainty.

\subsection[] {Surface brightness effects}

The total $\bj$ magnitudes used in this analysis have been derived
from plate scans (Maddox et al. 1990) and so they will be more susceptible
to certain systematic 
effects than those derived from CCD data.  In order to improve the
accuracy of our magnitudes several re-calibrations of the $\bj$
magnitudes have been made using CCD photometry from overlapping
fields (see Norberg et al., 2001).  

\begin{figure}
 \epsfig{file=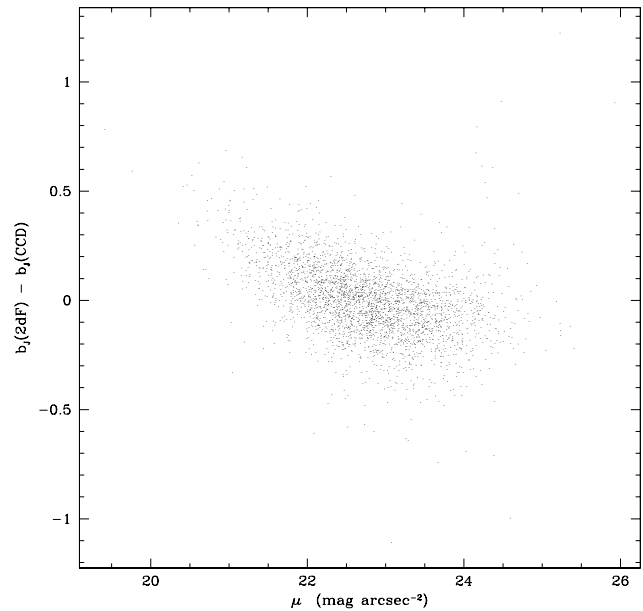,width=3.5in}
 \caption{The shift between the 2dFGRS $\bj$ magnitude and that derived
 from a CCD photometric survey (Cross et al., in preparation) is shown
 for $\sim3000$ galaxies.  It 
 can be seen that the shift varies systematically with surface
 brightness and so care must be taken when dividing the sample.}
 \label{surf1}
\end{figure}

Unfortunately some small
offsets still remain.  One such effect which may be important to the
analysis of LFs per spectral type is an apparent shift in $\bj$
magnitude with surface brightness (Cross 2001; Blanton et al.,
private communication).  Fig.~\ref{surf1} shows the offsets between the
2dFGRS $\bj$ magnitudes and those derived from a CCD photometric survey
(Cross 2001; Cross et al., in preparation) versus surface brightness,
a trend is clear in that 
high surface brightness objects tend to be systematically fainter in
2dFGRS $\bj$.  These offsets are most likely due
to a saturation effect in the (plate derived) 2dFGRS
magnitudes.  

\begin{figure}
 \epsfig{file=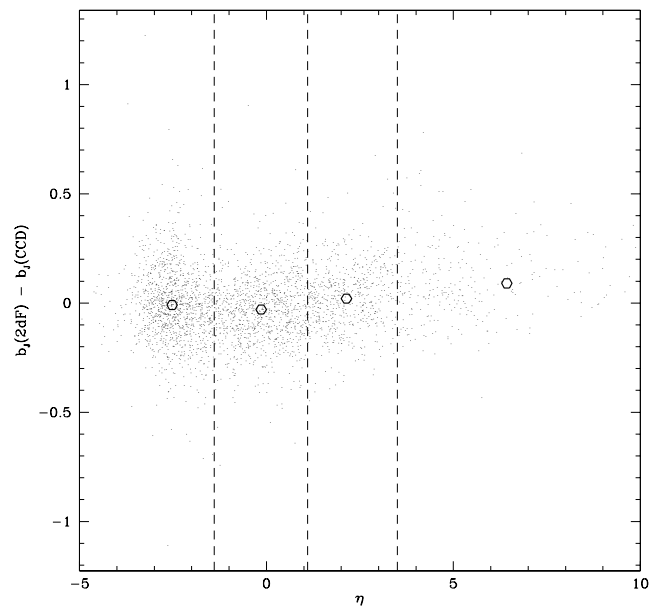,width=3.5in}
 \caption{The $\bj$ magnitude error versus spectral type ($\eta$) is shown.
 It can be seen that there is a weak correlation between the two
 quantities.} 
 \label{surf2}
\end{figure}

These small offsets will have a negligible effect on the
calculation of the overall LF but may be significant when one divides
the 2dFGRS sample in a way which is related to surface brightness.
Fig.~\ref{surf2} shows how these shifts depend upon our spectral
classification, it can be seen that the relationship is
surprisingly weak.  

The mean shifts we calculate per type are
however significant compared to their estimated uncertainties
(Table.~\ref{tabsb}), particularly for the most late-type galaxies.
We correct the galaxy magnitudes for each
spectral type in order to account for these shifts in the analysis
that follows.

\begin{table}
 \caption{The mean shifts between the 2dFGRS $\bj$ magnitudes and the
 CCD magnitudes per spectral type.  The stated uncertainties are $1\sigma$.}
 \begin{tabular}{@{}ccc@{}}
   \hline
   Type & Mean shift (mag) & Uncertainty \\
   \hline
     1 & $-$0.009  & 0.005 \\
     2 & $-$0.033  & 0.004 \\
     3 & 0.024    & 0.006 \\
     4 & 0.09     & 0.01 \\
   \hline
 \end{tabular}
 \label{tabsb}
\end{table}

\subsection[] {Estimators}

We assume a flat homogeneous Universe
with a uniform Hubble flow ($h=H_0/100$\,km\,s$^{-1}$\,Mpc$^{-1}$), a
cosmological constant 
$\Omega_\Lambda=0.7$ and mass density parameter $\Omega_{\rm{m}}=0.3$.  In
computing the LFs we use both the step-wise maximum likelihood (SWML) method
(Efstathiou, Ellis \& Peterson 1988, hereafter EEP88) to make a
non-parametric fit to 
the LF and the STY method (Sandage, Tammann \& Yahil 1979) to
calculate the maximum-likelihood Schechter function fit to the LF,
\begin{equation}
\phi(L)dL = \phi^*\left(\frac{L}{L^*}\right)^\alpha\exp\left(-\frac{L}{L^*}\right)\frac{dL}{L^*} \;.
\end{equation}
Both of these estimators are discussed extensively in the literature (e.g.
EEP88; Willmer 1997).
In addition we take into account the presence of Malmquist bias
in our STY fit by assuming that the errors in the $\bj$ magnitudes are Gaussian
($\sigma_M= 0.15$, see Colless et al. 2001). The observed LF will then
be a convolution of the true LF with the distribution of magnitude errors, 
\begin{equation}
\phi_{\rm obs}(M) =
\frac{1}{\sqrt{2\pi}\sigma_M}\int_{-\infty}^{\infty}\phi(M')e^{-(M'-M)^2
/2\sigma_M^2}dM' \;.
\end{equation}
Neither method predicts the normalisation of the LF so we 
estimate this independently by fitting to the projected counts,
$149.4/\sq\degr$ 
(Section~4.1). In so doing we make the assumption that the type
fractions in the parent catalogue are identical to those of the
classified galaxies (over the redshift interval $0.01\le z \le0.15$).

\section[]{Discussion}
\label{section:discussion}

\subsection{Results}

\begin{figure*}
 \epsfig{file=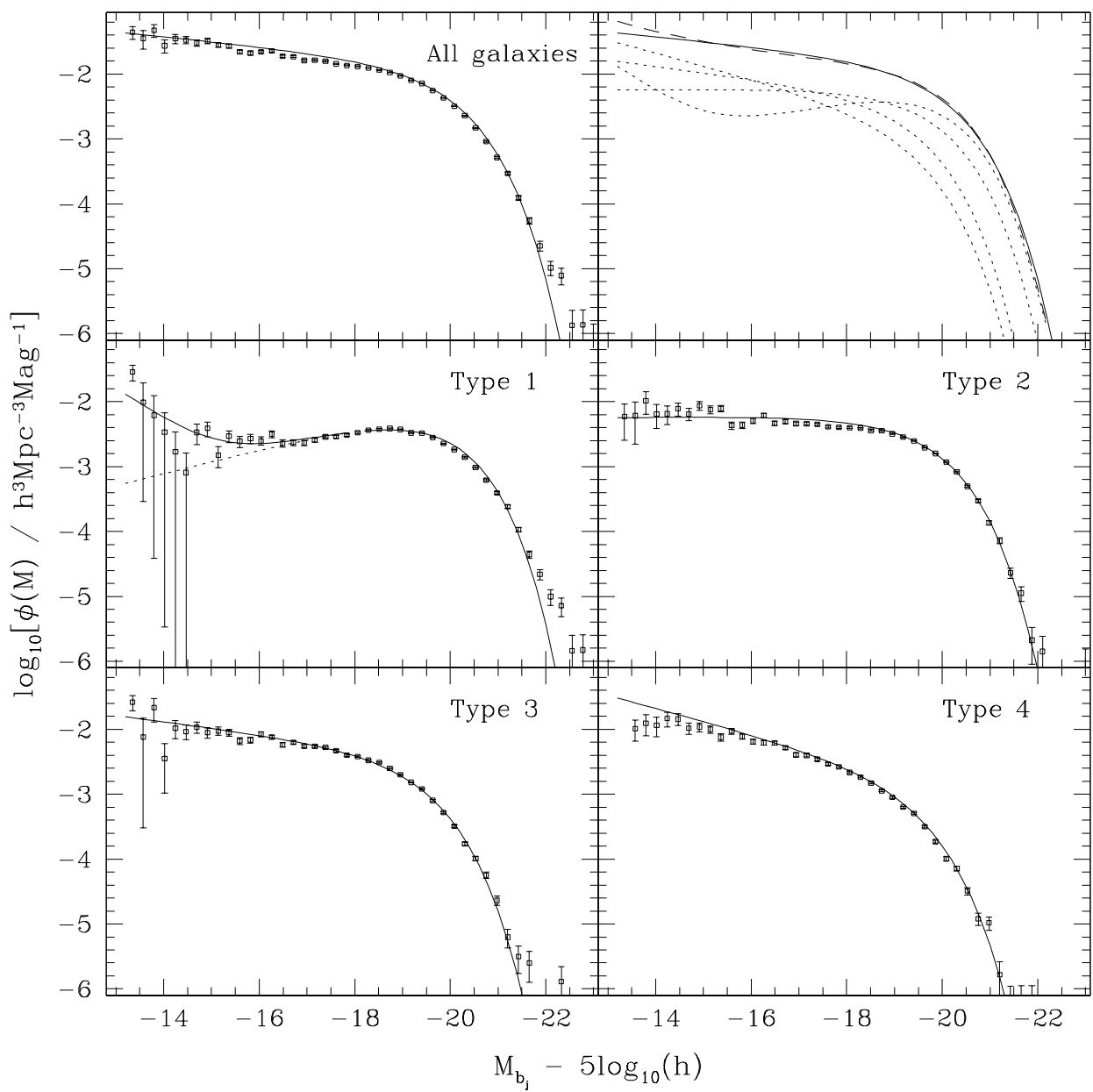,width=7in}
 \caption{The luminosity functions derived per spectral type.  The top
 left box shows the LF derived from all the classified galaxies, and
 the top right box shows how this (solid line) relates to the
 summation of the fitted LFs per type (dashed line). For each of the
 remaining four panels we show the SWML points and Schechter fits to
 the LF per spectral type.  The Schechter function has been fit over
 the entire range in $M_{\bj}$ magnitude for Types 2, 3 and 4, and only
 out to $M_{lim}-5\log_{10}(h)=-17$ for Type 1.  For Type 1 we show
 the best fitting Schechter function for the bright galaxies (dotted
 line) and the combination of this with our analytic form for the
 faint end residuals (solid line).  Our assumed cosmology here is
 $\Omega_{\rm{m}}$=0.3, $\Omega_\Lambda$=0.7.}
 \label{res}
\end{figure*}

Having performed the calculations outlined above we find that the
Schechter function provides an inadequate fit of the LF calculated
over the entire range in $M_{\bj}$ magnitude, especially for our most
passive and active star-forming galaxies (Types 1 and 4 respectively).
In order to determine a better fit to the LF of the passive
(Type 1) galaxies we have applied the following simple procedure:  
First we fit a Schechter function to the
LF of the bright galaxies ($M_{\bj}-5\log_{10}(h)<-17$) using the STY
method.  This provides an 
adequate representation of the data over the brightest ranges in
magnitude.  Having calculated this fit we then found that the residual
LF at the faint end could be well fit by a function of the form
\begin{equation}
\phi_{\rm res}(M_{\bj}) = 10^{[a + b M_{\bj}]} \;,
\end{equation}
particularly for the magnitude range $-17 <
M_{\bj}-5\log_{10}(h) < -14.5$ where the error-bars are small.
We calculated $a$ and $b$ using a least-squares fit to the
residual created by subtracting the STY fit from the SWML points at the
faint end.  The LF at the faint end can then be described by the combination
\begin{equation}
\phi(M) = \phi_{\rm{Sch}}(M;\alpha,M^*,\phi^*) + \phi_{\rm{res}}(M;a,b) \;,
\end{equation}
where $(a,b)=(4.7,0.50)$.  The uncertainties in these parameters are quite
large $(\pm0.9,\pm0.06)$, an obvious consequence of the large errors
in the SWML points at these magnitudes.

We estimate the scatter
in our derived Schechter parameters by performing a large number of random
realisations drawn 
from the estimated LFs. In addition to this sampling variance we
also take into account the uncertainties in our $K$-corrections
($20\%$) and in
the number density of the galaxies ($7\%$).  The quoted errors are
at the $1\sigma$ level. 

The two most notable features of our derived
LFs are the systematic variation of the faint end slope ($\alpha$)
and the characteristic magnitude ($M^*$) with spectral type.  The
faint end slope steepens significantly from $\alpha=-0.54$ for the
most passive star-forming galaxies (Type 1) to $\alpha=-1.50$ for the
most active galaxies (Type 4). On the other hand the characteristic
magnitude becomes systematically 
fainter as we go from the most passive ($M^*-5\log_{10}(h)= -19.58$)
to active 
($M^*-5\log_{10}(h) = -19.15$) star-forming galaxies. 
These results agree with previous measurements in that the late-type
galaxies are systematically fainter than their early-type counterparts.

Note that our results do
not change much if we exclude our correction for the errors in our
$\bj$ magnitudes
($\Delta M^*= 0.05-0.06$, $\Delta \alpha= 0.02-0.05$).

The fact that the Schechter function is not a good fit to the
data over the entire $M_{\bj}$ magnitude range is mostly due to an
over- (Type 1) or under- (Type 4)
abundance of faint objects relative to bright objects.  This is
particularly true of the passive (Type 1) sample of galaxies for which
there is a 
very significant increase in the predicted number density of faint
galaxies.  This feature has
been present in previous analyses (e.g. 
F99) however the small sample size has meant that only a statistically
insignificant number of galaxies have contributed.
Hence previous studies could not determine if this feature was real or
a consequence of the small
volume being sampled at these magnitudes.   For the first time here we
show significant evidence for the presence of a substantial dwarf
(passive star-forming) population with 142 galaxies having
$M_{\bj}-5\log_{10}(h)$ magnitude fainter than $-$16.0 ($\bar{z} = 0.027$). 
Of course these objects will tend to have low
signal-to-noise ratio spectra so some degree of contamination in this sample
should be present, particularly from erroneous redshift
determinations.  However, these preliminary results look promising and
we expect more detailed analyses to be forthcoming in the near future.

\begin{table*}
\begin{minipage}{160mm}
 \caption{Schechter function parameters derived in our analysis per
           spectral type using galaxies with $M_{\bj}<M_{\rm lim}$.
           The errors given are $\pm 1\sigma$.
           The normalisation, $\phi^*$, of the Schechter function is in
           units of $10^{-3}\,h^3$\,Mpc$^{-3}$ and
           $\rho_L=\phi^*L^* \Gamma(\alpha+2)$ is the luminosity
           density in units of $10^7L_{\sun} h\,$Mpc$^{-3}$
           ($M_{{\bj}{\sun}}=5.30$).} 
 \begin{tabular}{@{}ccccccccc@{}}
      \hline
  Sample & $\eta$  & $M_{\rm lim}-5\log_{10}(h)$ & \# Galaxies &
           $M^*-5\log_{10}(h)$ & $\alpha$ 
           & $\phi^*$ & $\rho_L$ \\
    \hline
       All   &  -  & $-13.0$ & 75589 & $-19.79\pm0.04$ & $-1.19\pm0.01$ &
           $15.9\pm1.0$ &  $19.9\pm1.4$ \\
 1 & $\eta<-1.4$ & $-17.0$ & 27540 & $-19.58\pm0.05$ &
           $-0.54\pm0.02$ & $9.9\pm0.5$ & $7.8\pm0.5$ \\
 2 & $-1.4\le\eta<1.1$ & $-13.0$ & 24256 & $-19.53\pm0.03$ & $-0.99\pm0.01$ &
           $7.2\pm0.4$ & $6.1\pm0.4$ \\
 3 & $1.1\le\eta<3.5$ & $-13.0$ & 15016 & $-19.17\pm0.04$ & $-1.24\pm0.02$ &
           $5.0\pm0.3$ & $3.7\pm0.3$\\
 4 & $\eta\ge3.5$ & $-13.0$ & 8386 & $-19.15\pm0.05$ & $-1.50\pm0.03$ &
           $2.4\pm0.2$ & $2.6\pm0.3$ \\
      \hline
 \end{tabular}
 \label{tabres}
\end{minipage}
\end{table*}

\subsection[]{Previous results}

\begin{figure}
 \epsfig{file=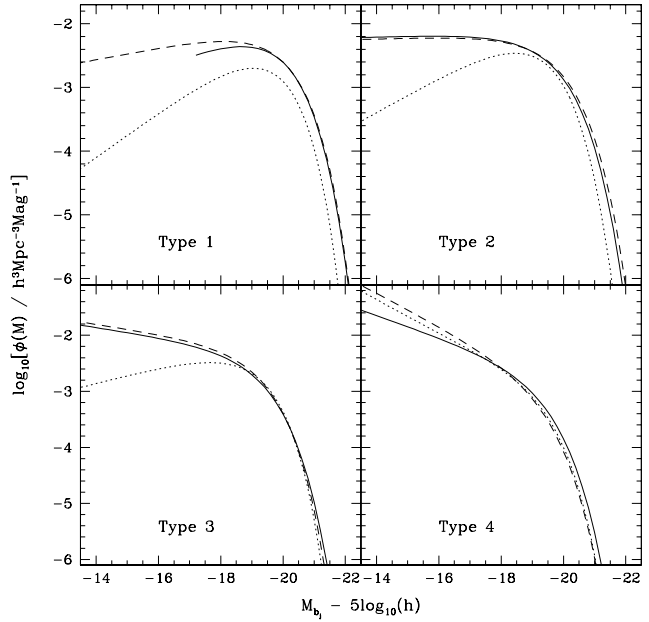,width=3.5in}
 \caption{Comparison with the type-dependent LFs calculated by
 Folkes et al. (dashed lines) and Bromley et al. (dotted lines).  The
 solid lines show our LFs assuming (for consistency) an Einstein
 de-Sitter Universe.  Note that the Type 1 LF has only been fit to
 $M_{\bj}-5\log_{10}(h) < -17$.  The
 disagreement between the different sets of determinations may be due
 to the poor fit of the Schechter function to the LF or to different
 selection effects between the surveys.}
 \label{cpsf}
\end{figure}

F99 has performed similar calculations of the LF per spectral type
using a subset of the data presented here.  In Fig.~\ref{cpsf} we
compare their results with those presented here.  Note that in order
to make this comparison we have recalculated the LFs assuming (for
consistency) an Einstein de-Sitter Universe ($\Omega=1$,
$\Lambda=0$), and we have reduced the 5 classes of F99 to
our 4 types by taking linear combinations of their LFs which
give the same fractions of galaxies per type as our own.
It can be seen from the figure that our results are in
good agreement with those of F99 except for some small systematic
shifts.  These differences are due to four factors
which we believe have significantly improved the accuracy of our
analysis.
\begin{enumerate}
\item We have used a considerably larger sample of galaxies in our
calculations. 
\item We have adopted self consistent $K$-corrections,
resulting in a small change in the predicted $M^*$ and $\phi^*$.
\item Taking into account recent advances in our CCD calibrations we
have been able to update the catalogue $\bj$ magnitudes (Colless et
al. 2001; Norberg et al., 2001).  This
has also resulted in a small change in the number counts and hence
$\phi^*$. 
\item We have corrected for saturation effects in high surface
brightness objects (Section~4.3).  This has a particularly large
impact on our most late-type (Type 4) galaxies.
\end{enumerate}

The most similar independent analysis of the LF per spectral type has
been conducted by Bromley et al. (1998), using the Las
Campanas Redshift Survey.  We also compare these results with our own
in Fig.~\ref{cpsf}.  Again, in order to compare the 6 clans with our 4
types we have had to combine their LF determinations in a way which
yields the same fractions of galaxies per type as our own.  In
addition we have applied the transformation $M_{\bj} = M_{\rm r}+1.1$ (Lin
et al. 1996) to convert the LCRS Gunn-$r$ Magnitudes to $\bj$.

The trend of the galaxy LFs to become progressively fainter as one
moves from passive to active star-forming galaxies is clear for both
sets of results.  However there appears to be some significant
disagreement particularly in the faint end slope between the
two LF determinations.  
Given the poor fit of the Schechter function to the faint end of the LF
we would not expect these two sets of results to necessarily agree
within their stated errors.  The large differences observed
are most probably due to the different selection effects
(e.g. surface brightness and $\bj$ versus $r$ selection) which can be 
particularly prominent for fainter
galaxies, hence adding more uncertainty to $\alpha$ than can be
estimated from a standard error analysis.

Our total luminosity density (again assuming an Einstein de-Sitter
Universe) is within 4\% of the figure obtained in F99 and is
30\% lower than the figure quoted by Blanton
et al. (2001) for the Sloan Digital Sky Survey early data release. The
reasons for this discrepancy are discussed in Norberg et al. (2001),
but in summary we believe that the Blanton et al. figure is
too high, partly owing to large-scale structure
in their small initial sample, and also through
the use of an inappropriate colour equation to predict 2dFGRS magnitudes.

\subsection[]{Aperture effects and evolution}

The spectra observed as part of the 2dFGRS are acquired through
optical fibres with a diameter of 140$\mu$m, corresponding to between $2''$ and
$2.16''$ on the sky (depending on plate-position, see Lewis et al.,
2001).  
In our assumed cosmology ($\Omega_{\rm{m}}$=0.3, $\Omega_\Lambda$=0.7) 
this translates to an aperture width of
approximately $2.8h^{-1}$~kpc at the survey median redshift of
$z=0.1$.
The fact that fitted galaxy brightness profiles have scale lengths 
of the order of 3-4 kpc (Binney \& Merrifield 1998) raises the
question of how representative the observed spectra are of the galaxies
as a whole.  However, with typical
seeing ($1.5''-1.8''$) and random fibre positioning errors (RMS $0.3''$) 
the  2dF fibre-aperture will in fact sample a much larger portion of
each galaxy.

If the effective fibre apertures were significantly smaller than
the galaxies over a range of redshifts then the variation in 
this aperture with redshift would result in systematic biases in the
determination of the LFs.  These so-called `aperture effects' are difficult
to quantify without an in depth knowledge of the spatially resolved
spectral properties of the targeted galaxies. 

\begin{figure}
 \epsfig{file=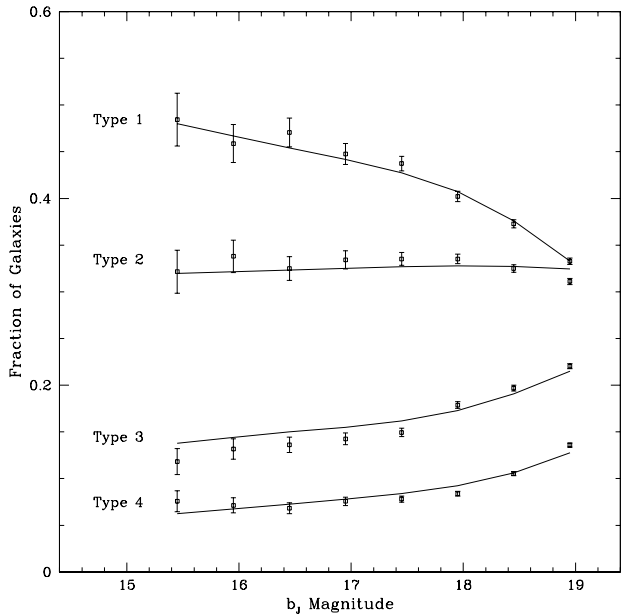,width=3.5in}
 \caption{A test recently proposed by Kochanek, Pahre \& Falco (2000).
 We show 
 observed (squares) and predicted (lines) type fractions for
 the 2dF galaxies used in our calculations, where the predicted fractions have
 been derived from integrating our LFs. The two are consistent within the
 stated (Poisson) errors.}
 \label{aper}
\end{figure}

Kochanek, Pahre \& Falco (2000) recently suggested a test to estimate the
influence of this fibre-aperture
bias on the determination of type dependent LFs $\phi_i(M)$.  The
test involves comparing the 
observed fractional number counts of galaxies in each 
class to the fractional number counts predicted by our derived LFs.
Kochanek, Pahre \& Falco found that 
for redshift surveys classified morphologically the two were generally
consistent but that for surveys classified spectrally using
fibre-aperture spectra the two were very inconsistent.  We have performed
this test on our data and found no substantial inconsistencies
(Fig.~\ref{aper}).  With the exception of small systematic offsets the
observed type fractions were found to agree with those predicted by
the LF to within 2$\sigma$ over the entire range in $\bj$ magnitude.
We note that as the number counts predicted by the LF are normalised
by construction to the observed number counts this test is only
necessary and not sufficient.  For this reason more sophisticated
tests such as comparisons with detailed models
are necessary in order to isolate and analyse the competing effects on
the data (e.g. positional errors, seeing and evolution).

\begin{figure}
 \epsfig{file=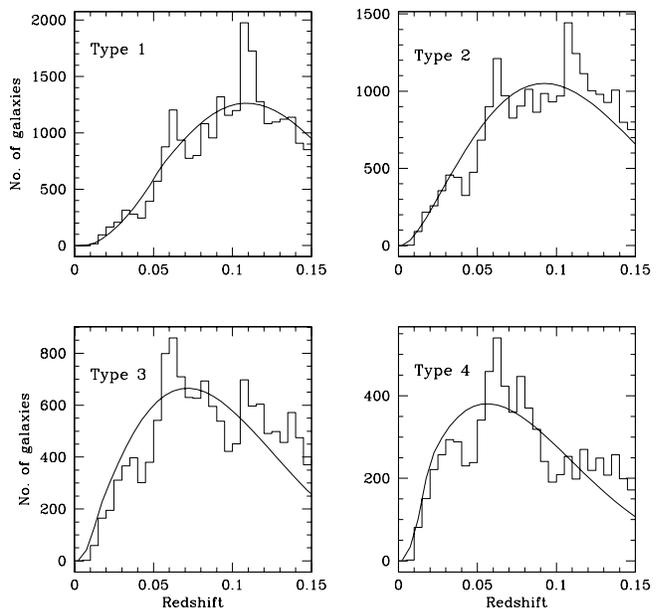,width=3.5in}
 \caption{Comparison between the observed $N(z)$ distributions
 (histograms) and the predicted $N(z)$ for each spectral type
 ($\bj<19.3$).  Note
 that the $N(z)$ distributions of the first two types are recovered
 very well but that significant deviations exist for the last two
 (most active star-forming) types.}
 \label{nofzfig}
\end{figure}

We show the observed and predicted $N(z)$ distributions for each
spectral type out to $\bj<19.3$ in Fig.~\ref{nofzfig}.  
It can be seen that the observed
$N(z)$ is well recovered by the LF for the first two types.
There appears to be a significant over
abundance of high-$z$ galaxies for the two highest star-forming types
as well as a slight under abundance of low-$z$ galaxies.  Similar
deviations in abundance could be explained by aperture effects.
However, if this were the only explanation then we would also observe
a significant over abundance of low-$z$ passive star-forming galaxies,
which is not the case.  A much more reasonable explanation for these
deviations between 
the predicted and observed $N(z)$ distributions is the presence of
evolution (see e.g. Lin et al. 1999).  Because these two effects affect
the observed galaxy population in similar ways they will be difficult
to dis-entangle in more detailed analyses.  For the present
we note that as far as the determination of the LF in broad bins of
spectral type is concerned both of these biases have only a small
effect on our calculations.

\section{Conclusions}
\label{section:conclusions}

In this paper we have applied a method of spectral classification,
$\eta$, based on the the relative emission/absorption line strength
present in a galaxy's optical spectrum.  This classification
correlates well with morphology and we interpret it as a measure of
the relative current star-formation present in each galaxy.  Based on
this classification we have divided the galaxies observed to date in
the 2dFGRS into 4 spectral types and calculated the LF for each.
Our results are consistent with previous determinations in that the
faint end slope systematically steepens and the characteristic
magnitude becomes fainter as we move from passive to active star-forming galaxies.  
In addition we have been able to show that the Schechter function is a
poor fit to the LF over a large range of $M_{\bj}$ magnitudes, perhaps
as a result of sub-populations within the data.  The
calculations presented in this letter represent the most accurate
determinations of the optical $\bj$ galaxy LF to date.

As $\eta$ is a continuous variable the next step in quantifying the 
variation of the LF with spectral
type is to develop a bivariate formalism $\phi(M,\eta)$.  Such a
result would be extremely valuable in that it would quantify the
variation in the Schechter parameters with galaxy type, $\alpha(\eta)$ and
$M^*(\eta)$, allowing the 
results to be applied much more objectively to such things as
semi-analytic models 
from which spectra as well as other galaxy properties can be derived
(e.g. Slonim et al. 2001 and references therein).  We can then also
extend this  
approach of comparing the observed LFs with model LFs by mimicking the 
observational selection effects (e.g. aperture bias).

One of the key advantages of our spectral type, $\eta$, is that it can
easily be applied to any other set of galaxy spectra which occupy the
relevant wavelength range.
This will be invaluable when
comparing results from the many different galaxy surveys currently underway 
(e.g. SDSS, VIRMOS, 6dF etc.).

\bigskip

The 2dF Galaxy Redshift Survey has made available to the community 100,000
galaxy spectra and redshifts through their website,
{\tt{http://www.mso.anu.edu.au/2dFGRS/}}.  Additional information
regarding the spectral classification and LFs presented here can be
obtained from {\tt http://www.ast.cam.ac.uk/$\sim$twodfgrs/}.

\section*{Acknowledgments}

DSM was supported by an Isaac Newton Studentship from the University of 
Cambridge and Trinity College.  We thank R. Kaldare, S. Ronen and R. 
Somerville for many helpful discussions.  The 2dF Galaxy Redshift Survey 
was made possible through the dedicated efforts of the staff at the Anglo Australian Observatory, both in creating the two-degree field instrument and supporting it on the telescope.

\bsp
\label{lastpage}
\end{document}